# HUBBLE SPACE TELESCOPE OBSERVATIONS
# OF THE LENSING CLUSTER ABELL 2218

J.-P. Kneib,[1] R.S. Ellis,[1] I. Smail,[2] W.J. Couch[3] & R.M. Sharples[4]

1) Institute of Astronomy, Madingley Road, Cambridge CB3 0HA, U.K.

2) The Observatories of the Carnegie Institution of Washington, 813 Santa Barbara St.,
Pasadena, CA 91101-1292

3) School of Physics, University of NSW, Sydney 2052, NSW Australia

4) Dept. of Physics, University of Durham, South Road, Durham DH1 3LE, U.K.

## ABSTRACT

We present a striking new Hubble Space Telescope (HST) observation of the rich cluster Abell 2218 taken with the Wide-Field and Planetary Camera (WFPC2). HST's restored image quality reveals a sizeable number of gravitationally-lensed features in this cluster, significantly more than had been identified using ground-based telescopes. The brightest arcs are resolved by HST and show internal features enabling us to identify multiply-imaged examples, confirming and improving the mass models of the cluster determined from ground-based observations. Although weak lensing has been detected statistically in this and other clusters from ground-based data, the superlative resolution of HST enables us to individually identify weakly distorted images more reliably than hitherto, with important consequences for their redshift determination. Using an improved mass model for the cluster calibrated with available spectroscopy for the brightest arcs, we demonstrate how inversion of the lensing model can be used to yield the redshift distribution of $\sim$80 faint arclets to $R \simeq 25$. We present a new formalism for estimating the uncertainties in this inversion method and review prospects for interpreting our results and verifying the predicted redshifts.

*Subject headings:* cosmology: observations – galaxies: evolution – gravitational lensing

## 1. Introduction

The gravitational lensing of faint background galaxies by rich clusters is emerging as a very promising method to constrain both the distribution of dark matter in clusters and the statistical redshift distribution of galaxies beyond the reach of conventional spectrographs (Fort & Mellier 1994). The lensing distortion induced in the image of a typical distant galaxy by a foreground rich cluster depends upon the product of a scale factor (involving the galaxy and cluster redshifts and the adopted cosmological model) and the second derivatives of the projected cluster potential. The majority of the faint lensed images are only weakly distorted and these are termed "arclets". However, a small fraction are highly distorted "giant arcs" – images which lie near critical lines and suffer high amplification; these are particularly helpful in mass modelling since their relatively bright magnitudes mean that they can often be studied spectroscopically. With redshifts for one or more giant arcs in a cluster the absolute mass of the central regions can be accurately determined. Multiply-imaged sources, even without redshifts, provide additional information on the geometrical configuration of the potential well in the core regions (Mellier et al. 1993, Smail et al. 1995a). Recent work has concentrated on clusters with arcs of known redshift *and* multiply-imaged sources. In such cases, a robust model of the cluster mass can be constructed, allowing inversion of the lens equations for the arclet population and yielding the redshift distribution of extremely faint galaxies.

For the well-studied cluster Abell 370 ($z_{cl} = 0.37$), Kneib et al. (1994a) demonstrated a first application of this inversion technique by identifying $\sim 30$ candidate arclets with axial ratios $a/b \gtrsim 1.4$ from ground-based images taken in superlative conditions. For each arclet, unlensed magnitudes and probable redshifts to a limit of $B \simeq 27$ were inferred from a detailed mass model calibrated by the redshift of a giant arc and the properties of various multiple images (Kneib et al. 1993). However, this new technique suffered from several uncertainties. Firstly, simple mass models may ignore substructure in the cluster mass distribution leading to imprecise inversion. Also, in the absence of spectroscopic or morphological data, some of the candidate multiply-imaged objects used to model the form of the potential may be spuriously identified in ground-based data.

Finally, even in the best ground-based conditions, the limited angular resolution makes it difficult to distinguish lensed arclets from intrinsically-elongated faint sources and to accurately measure their shapes; such confusion may lead to contamination of the inverted redshift distribution by cluster members, foreground spirals, and close galaxy pairs.

Even in its aberrated state the advantages of HST for lensing studies over the best ground-based telescopes soon became evident (Smail et al. 1995b). Here we illustrate that the *refurbished* HST is even more powerful, allowing reliable identification of multiple images and faint arclets. Considerable progress is thus possible with HST in the inversion method developed by Kneib et al. (1994a).

A plan of the paper follows. Section 2 describes the observations and gives a qualitative description of the HST images, including those lensed features which allow us to improve the ground-based model of Kneib et al. (1995). Section 3 describes the improved mass model we have implemented. Starting from the mass model of a cluster, Section 4 introduces the theory of the lens inversion and discuss the probability distribution of the redshift of a sheared galaxy. The sources of uncertainty in this inversion are also discussed in the context of observational data. In Section 5 we present our results on the faint field galaxy redshift distribution and discusses both the limitations of comparing such results with model predictions as well as the prospects for verifying the inverted redshifts with further observations. Section 6 summarises the overall conclusions of the paper. Throughout this paper, we assume $H_0 = 50$ km s$^{-1}$ Mpc$^{-1}$, $\Omega_0 = 1$ and $\Lambda = 0$.

## 2. Observations and Analysis

### 2.1. Previous Observations of Abell 2218

Abell 2218 (Fig. 1a), is one of the best-studied rich clusters at intermediate redshift ($z_{cl} = 0.175$). Le Borgne et al. (1992) present a detailed photometric and spectroscopic survey of the cluster and derive a rest-frame velocity dispersion of $\sigma_{cl} = 1370^{+160}_{-120}$ km s$^{-1}$, indicative of a deep potential well. This is supported by a high X-ray luminosity ($L_x(0.5\text{-}4.4\ \text{keV}) = 6.5 \times 10^{44}$ ergs s$^{-1}$) and a strong Sunyaev-Zeldovich decrement (Jones et al. 1993, Birkinshaw & Hughes 1994). The cluster contains a number of luminous giant arcs, discovered and extensively studied by Pelló et al. (1988, 1992). Several of the brighter arcs have been observed spectroscopically; redshifts for these and ground-based colors for other lensed features provide the basic ingredients for the recent mass model of Kneib et al. (1995). Using four systems of arcs and possible counter-arcs, tentatively identified from ground-based colors (#289, #359-#328, #384-#468 and #730 in the numbering scheme of Le Borgne et al. (1992)), Kneib et al. (1995) determine a mass distribution for the cluster core which is bimodal in form and concentrated around the two most luminous cluster galaxies (#391, #244, Table 3).

## 2.2. HST Observations and Photometric Catalogue

Abell 2218 was observed by the HST WFPC-2 camera on September 2, 1994. Three exposures totalling 6500 sec were taken through the F702W filter. Each exposure was shifted relative to the others by 3 WFC pixels (0.30 arcsec) providing a partial overlap of the chip fields. After pipeline processing, standard IRAF/STSDAS routines were employed to shift and combine the frames to remove both cosmic rays and hot pixels. We discard the PC chip from our analysis because of its brighter isophotal limit. The final frame comprising the 3 WFC chips (Fig. 1a & b) has an effective resolution of 0.14 arcsec and a $1\sigma$ detection limit per resolution element of $R \simeq 30$. We convert our instrumental F702W magnitudes into standard $R$ using the synthetic zero point and color corrections listed in Holtzman et al. (1995). For the color term we choose $(V - R) \simeq 0.6$ typical of the faint field population (Smail et al. 1995d). The color correction is +0.2 mag, and the typical photometric errors of our faintest objects, $R < 25.5$, are $\delta R \sim 0.1$–0.2.

To produce a catalogue of faint arclets from our data we first analysed the HST image using the Sextractor package (Bertin 1995, Bertin & Arnouts 1995). All objects with isophotal areas above 12 pixels (0.12 arcsec$^2$) at the $\mu_R = 24.8$ mag arcsec$^{-2}$ isophote (2$\sigma$/pixel) were selected. A comparison of the differential number counts of these images to deep ground-based $R$ counts (Smail et al. 1995d) shows a marked excess of galaxies brighter than $R \sim 21.5$ due to cluster members (Figure 2), and a steep roll-over in the observed counts beyond $R \sim 25$ arising from an incompleteness which amounts to 55% in the $R = 25$–26 bin. We thus applied a magnitude limit of $R = 26$ giving a total of 440 images over a 4.7 sq. arcmin area. A neural-network algorithm (Bertin & Arnouts 1995) was used to separate stars and galaxies leading to the exclusion of 25 star-like objects from the catalogue.

From this list we selected a sample of candidate arclets, first removing all galaxies with $R < 21.5$ (probable cluster members) and objects lying in the halos of giant ellipticals and very faint object $R > 25$ as their photometry and shapes are uncertain. The procedure reduced our catalog to $\sim 235$ arclet candidates.

### 2.3. Multiply-Imaged Features

At this stage, it is useful to review the multiply-imaged features identified on the HST image in the context of the ground-based predictions, prior to using them to improve the mass model of Kneib et al. (1995).

Four bright arcs and counter-arcs were identified as matching images by Kneib et al. (1995) on the basis of their ground-based colours. Each of these is clearly resolved by HST with internal structures that enable us to verify their multiply-imaged nature (see Fig. 1c of Smail et al. 1995b). We discuss each of these images here and summarise their photometric properties in Table 1.

**#384 and #468** #384 is a most impressive arc system with an internally-symmetric pattern of unresolved knots showing that this image is clearly formed from the merger of two images of reversed parity. This enables the location of the critical line to be accurately identified. The knots, which presumably represent HII regions in a blue star-forming galaxy, can also be seen in the counter image #468. A further feature of interest is the train-track-like morphology of the source, also replicated in #468.

**#359, #328, #337 and #389** The red arc #359 has a spectroscopic redshift of $z = 0.702$ and

shows no internal structure; this is consistent with its identification as a background spheroidal galaxy. It was naturally interpreted as a fold arc i.e. two merging images (Kneib et al. 1995) with a single counter-image #328. The absence of a strong discontinuity (even in the HST image) in the surface brightness along the #359 arc can be explained if the surface brightness peak lies just outside (or on) the caustic on the source plane. However, a detailed inspection of the HST image demonstrates that this simple picture is unlikely to be correct as a faint extension of #359 is now revealed which merges with #337. From the ground-based data it was noticed that #337 & #389 had similar colours to #359 but no simple model was able to explain such a configuration.

If #337 is indeed a counter-image of arc #359, then we may consider whether #389 is also a counter-image. In Section 3, we will show that, by incorporating individual cluster galaxies in the mass model, it is straightforward to show that #359 is a fold-arc with #328, #337 and #389 each as counter-images.

**#289** In contrast to #359, the blue arc #289 with a spectroscopic redshift of $z = 1.034$ exhibits a large amount of internal structure. Although the arc is luminous and therefore probably highly magnified, the bright southern end does not appear particularly strongly sheared and is apparently not multiply-imaged. Close inspection of the northern section of this arc indicates that it extends across the halo of the cluster galaxy #244. The complex morphology can thus be explained via a background galaxy straddling the caustic. The majority of the source lies outside the caustic producing a single highly magnified, weakly sheared image. The portion within the caustic is multiply-imaged and produces the highly elongated tail across the halo of #244.

A detailed examination of the HST image reveals several new potentially-important multiply-imaged systems.

**#730** This very faint thin arc was suggested as a possible lensed feature in the ground-based data but is clearly verified as such by HST. The faintness makes it difficult to identify the individual sub-components at this stage, although a number of bright knots are visible. Nevertheless, the structure suggests a likely cusp arc as three components can be distinguished.

**H1–3, H4–5** These are two impressive multiply-imaged systems which were unrecognised in the ground-based studies (Fig. 1b). From the morphologies and positions, H1–3 appear to be three images of a section of the disk of #273, the remainder of the source being only singly-imaged. The very faint features H4–5 ( R≈26) are believed to represent a new very faint multiply-imaged pair. Several candidates for the counter-image to this pair exist on the opposite side of the cluster.

**#444 + H6.** #444 is a fold arc (two merging images) with H6 as a counter image.

In summary, the HST image not only allows us to confirm the lensed features that underpin the ground-based mass model, but also provides additional information that enables us to refine the model. We have identified a total of 7 multiply-imaged sources seen through the core of Abell 2218. This is a substantial improvement over the ground-based tally and significantly more than the number known in any other cluster at this time. By analysing these features we can thus hope for the most detailed view of the mass distribution within a cluster thus obtained. The model refinements derived from these new multiply-imaged features are principally in the detailed form of the mass model and lead to little change in the global cluster mass/light ratio. However, they can have an effect on the lensing inversion and we will explore this further in Section 5.

### 2.4. Arclets and Shear

Using our previously-defined catalogue of faint sources ( §2.2), we now construct a "shear (or deformation) map" defined as the local average of the deformation vector (see §4) of the lensed galaxies:

$$< \overline{\tau} > (x, y) = \int \int \overline{\tau}_I(x', y') \omega(x - x', y - y') dx' dy', \tag{1}$$

where $\omega(x, y)$ is a normalized weighting function. The weighting function chosen was a gaussian of 20 arcseconds FWHM. Fig. 4 shows the deformation map within the field of the WFC upon which we have superimposed the location of some of the most luminous cluster members. This map provides a view of the cluster potential with a resolution of 20 arcsecs ($\sim$ 75kpc), allowing us to detect any substructure in the cluster mass distribution on scales larger than the resolution. We now use this information to assist in the construction of a refined model for Abell 2218 taking into account the new details on the multiple images discussed in §2.3.

It is important to recognise that this map differs from the true deformation for a number of reasons including contamination by foreground and cluster galaxies and also the dependency of the strength of the image deformation on redshift. The former effect has already been minimised by removing the brightest galaxies from our list. Furthermore, we can assume that the bulk of the contaminating galaxies are randomly orientated and thus only dilute the modulus of the shear, without affecting its form. The earlier ground-based mass model was bimodal in form and centred on the cD (#391) and #244. The shear map in these regions suggests contributions from #196 and #235 should now be included.

Although the shear map is statistical in nature, the HST resolution has encouraged us to define a visually-selected sample of the brighter and larger arclets, intermediate to the bright arcs reviewed in the previous section. These sources are sheared sufficiently that their identification as lensed features is in little doubt and furthermore most of them are within spectroscopic reach. Their properties are summarised in Table 2 and Fig. 3 compares their distribution with the 20 "arclets" identified from ground-based data (Pelló et al. 1992). The total number of HST arclets is now significantly increased. Furthermore, the HST data suggests that as many as a third of the ground-based arclets are close galaxy pairs or misidentified edge-on disk galaxies.

In summary, the improved resolution of the repaired HST allows considerable progress to be made in the identification and understanding of lensed features in Abell 2218. The resolution of the brighter arcs confirms several of the multiply-imaged features suggested from the ground-based studies. In particular, the fold arc #359 is now identified as a 5-image configuration, and a number of new multiply-imaged candidates are revealed. Similarly, the HST image allows us to identify weakly-lensed features (arclets) with greater reliability, both on an individual basis and statistically. In both respects, we are better placed to refine the mass model developed on the basis of ground-based imaging and to identify arclets for redshift determination.

## 3. Mass Modelling

The mass modelling method we use is based on the precepts developed by Kneib (1993) which has now been successfully applied to describe many different cluster lenses including MS2137 (Mellier et al. 1993), A370 (Kneib et al. 1993), Cl2236 (Kneib et al. 1994b), Abell 2218 (Kneib et al. 1995) and Cl0024 (Smail et al. 1995b).

The basic approach is to use multiply-imaged systems and the mean orientation of the arclets to constrain an analytical representation of the total mass based upon components associated with likely centres of mass, i.e. massive cluster galaxies. Each component is described by a minimal set of parameters: position, ellipticity, orientation, core size and central velocity dispersion. The associated mass distribution should be approximately isothermal if the central mass is relaxed. The particular analytical expression used is based on the pseudo-isothermal elliptical mass distribution (PIEMD) with ellipticity $e = (a - b)/(a + b)$ derived by Kassiola & Kovner (1993):

$$\Sigma(x, y) = \Sigma_0 \frac{r_c}{\sqrt{r_c^2 + \rho^2}} = \frac{\sigma_0^2}{2G\sqrt{r_c^2 + \rho^2}} \ , \tag{2}$$

with

$$\rho^2 = \frac{x^2}{(1 + \epsilon)^2} + \frac{y^2}{(1 - \epsilon)^2} \ . \tag{3}$$

This expression has the advantage of describing mass distributions with arbitrarily large ellipticities. For each component used, we smoothly truncate the elliptical mass distributions (c.f. appendix of Kassiola & Kovner, 1993) using a linear combination of 2 PIEMD components:

$$\Sigma(x, y) = \Sigma_0 \frac{r_c r_{cut}}{r_{cut} - r_c} \left( \frac{1}{\sqrt{r_c^2 + \rho^2}} - \frac{1}{\sqrt{r_{cut}^2 + \rho^2}} \right) \ , \tag{4}$$

where $r_{cut}$ is the truncation radius (the surface mass density falls as $r^{-3}$ for $r >> r_{cut}$). The total mass of such a truncated mass distribution is finite and for $r >> r_{cut}$ in the limit where $e \to 0$:

$$M_{tot} = 2\pi \Sigma_0 r_c r_{cut} = \frac{\pi}{G} \sigma_0^2 r_{cut}^2 \ . \tag{5}$$

The ground-based mass model for Abell 2218 (Kneib et al. 1995) was based on two major components associated with galaxies #391 and #244. As discussed in Section 2.3, the detailed information now available from the multiple-images (particularly #359 and its counter-images) and the fine structure visible in the shear map (Fig. 4), encourages us to improve on this model by incorporating the effect of halos associated with components around #235 and #196 and other individual cluster galaxies. For each component the center, ellipticity and orientation are matched to those observed for the associated light distribution (as is the case for those associated with #391 and #244). However, the dynamical parameters $r_c, r_{cut}$ and $\sigma_0$ for these four main components are kept as free parameters.

When including galaxy-scale components into our model it is clear that such a refinement could, in principle, be continued indefinitely. In practice, we included all galaxies with R<19.5 (as the magnitude increases the mass of each galaxy become small and their lensing effects become negligible). In total, we incorporate halos associated with 30 luminous cluster galaxies into the mass model. For each halo the ellipticity and orientation match those observed for the galaxy light distribution. The other mass parameters are scaled according to the galaxy luminosity following Brainerd, Blandford & Smail (1995).

$$\sigma_0 = \sigma^* (L/L^*)^{1/4} \tag{6}$$

and

$$r_{cut} = r_{cut}^* (L/L^*)^{1/2} \tag{7}$$

where $\sigma^*$ and $r_{cut}^*$ are free parameters in the minimization procedure. Furthermore to have a profile that is identical from one galaxy to another we scale the core radius $r_0$ in the same way as $r_{cut}$:

$$r_0 = r_0^* (L/L^*)^{1/2} \tag{8}$$

The mass of individual galaxy scale as the luminosity with:

$$M_{tot} = \frac{\pi}{G} (\sigma^*)^2 r_{cut}^* (L/L^*) \tag{9}$$

It is worth emphasising that, by themselves, the individual galaxy halos do not contain enough mass to reproduce all of the lensed features observed in the cluster. In other words we must retain cluster-scale mass components associated with the brighter cluster galaxies (the central cD (#391), #235, #196 and #244).

To constrain the composite mass model we first define a $\chi^2$ estimator as the quadratic sum of the differences between the source parameters (position, orientation and ellipticity) for each set of multiple images (see Table 1), plus the observed shear as represented by the quadratic sum of $\tau_{pot} < \tau_I > \sin(2(\theta_{pot} - <\theta_I>))$. We then minimise this estimator by varying the parameters of the mass model. To stabilise and speed up the convergence we specify the location of the infinite magnification point in the fold or cusp images (i.e. the location of the symmetry break in the case of #384 or the luminosity peak of #359 and the saddle between #359 and #337 in the case of the arc #359 at z=0.702).

The best fiducial model resulting from the HST data is presented in Table 3. A contour plot of the mass distribution is shown in Fig. 5 where the shear-field is also shown for a source plane at $z_S = 1$. This mass model is currently the most detailed derived for a cluster core, the detail only being possible because of the combination of the high resolution shear map and the large number of multiply-imaged sources.

Although the difference between the HST and ground-based mass models is small when considering global properties such as cluster mass/light ratio, we show in Section 5 that there can be variations in the lensing inversion for specific arclets, depending upon their location. The principle change is in the detailed granularity of the mass distribution leading to a more precise inversion.

## 4. Gravitational Lensing Formalism

We now turn to the primary purpose of the paper, namely to take our well-constrained mass model for Abell 2218 and use it to derive statistical redshift distributions for the large sample of faint arclets discussed in §2.4. In what follows we extend the original discussion of Kneib et al. (1994a), developing a formalism for estimating the errors in the inversion redshifts of individual galaxies. This will be particularly useful as we have a range of lensed features from relatively bright arclets, many of which can be recognised as lensed on an individual basis, to fainter images which can only be treated statistically.

## 4.1. General Equations

The gravitational lensing formalism we use is based on the original treatise presented by Kneib et al. (1994a). The lens mapping is described by the transformation:

$$\vec{u_S} = \vec{u_I} - \mathcal{D}\vec{\nabla}\phi(\vec{u_I}) \ , \qquad (10)$$

where $\vec{u_S}$ is the position of the source, $\vec{u_I}$ is the position of the image, $\mathcal{D}$ is the dimensionless ratio $D_{LS}/D_{OS}$ and $\phi$ is the projected Newtonian potential normalized by $2/c^2$.

A distant galaxy can be described to the first order by five geometrical parameters: its centroid $(x_c, y_c)$, complex deformation $\overline{\tau} = \tau \ e^{2i\theta}$ and size $s$.

The first moment of the weighed surface brightness $\mu(x, y)$ distribution gives the position of the centroid $(x_c, y_c)$:

$$x_c = \frac{1}{\mu_W} \int \int W(x,y) \ \mu(x,y) x dx dy \quad y_c = \frac{1}{\mu_W} \int \int W(x,y) \ \mu(x,y) y dx dy \ , \qquad (11)$$

with

$$\mu_W = \int \int W(x,y) \ \mu(x,y) dx dy \ . \qquad (12)$$

The weighting function $W(x, y)$ can be adjusted to minimise the error in the determination of the centroid.

The second order moment matrix $M$ gives the shape of the galaxy ($\overline{\tau} = \tau \ e^{2i\theta}$), i.e. its equivalent ellipse of major-axis $a$, minor-axis $b$ and orientation $\theta$:

$$M = \frac{1}{\mu_W} \int \int W(x,y) \ \mu(x,y) x_i x_j dx dy = \begin{pmatrix} M_{xx} & M_{xy} \\ M_{xy} & M_{yy} \end{pmatrix} \propto R_\theta \begin{pmatrix} a^2 & 0 \\ 0 & b^2 \end{pmatrix} R_{-\theta} \ , \qquad (13)$$

where $R_\theta$ is the rotation matrix of angle $\theta$. Note that different weighting functions can be chosen in computing the first and second moment integrals, depending upon which is required with higher accuracy. The weighting factor is more critical in dealing with ground-based data than with HST images because of the effects of seeing (Bonnet & Mellier 1995, Kaiser et al. 1995, Wilson, Cole & Frenk 1995). In our analysis we used the simple weighting function:

$$W(x, y) = \begin{cases} 1 & \text{if } \mu < \mu_{ISO} \\ 0 & \text{if } \mu > \mu_{ISO} \end{cases} \tag{14}$$

The size parameter ($s$) is defined as:

$$s = 2\sqrt{\det M} \propto 2ab. \tag{15}$$

and the deformation matrix $D$ is:

$$D = \frac{M}{2\sqrt{\det M}} = \begin{pmatrix} \delta + \tau_x & \tau_y \\ \tau_y & \delta - \tau_x \end{pmatrix} , \tag{16}$$

where $\overline{\tau} = \tau_x + i\tau_y = \tau \, e^{2i\theta}$ is the complex deformation and $\delta = \sqrt{1 + \tau^2}$ is the real distortion parameter. In terms of the major and minor axis these are:

$$\tau = \frac{a^2 - b^2}{2ab}, \quad \delta = \frac{a^2 + b^2}{2ab}. \tag{17}$$

Further, the complex shear $\overline{g}$ and the complex ellipticity $\overline{\varepsilon}$ are defined as:

$$\delta = 1 + \overline{g} \, \overline{\tau}^* , \quad g = \frac{a - b}{a + b} , \tag{18}$$

and

$$\overline{\varepsilon} = \frac{\overline{\tau}}{\delta} , \quad \varepsilon = \frac{a^2 - b^2}{a^2 + b^2} , \tag{19}$$

($*$ denotes the conjugate of a complex number).

The lensing equation for the moment matrix is given by (Kochanek, 1990):

$$M_S = a^{-1} M_I \, {}^t a^{-1} , \tag{20}$$

where the subscript $S$ refers to the source, $I$ to the image and $a^{-1}$ is the inverse of the amplification matrix, defined as the Hessian of the lens mapping (eq. 10):

$$a^{-1} = \begin{pmatrix} 1 - \mathcal{D}\partial_{xx}\phi & -\mathcal{D}\partial_{xy}\phi \\ -\mathcal{D}\partial_{xy}\phi & 1 - \mathcal{D}\partial_{yy}\phi \end{pmatrix} \equiv \begin{pmatrix} 1 - \kappa + \gamma_x & \gamma_y \\ \gamma_y & 1 - \kappa - \gamma_x \end{pmatrix} . \tag{21}$$

$\kappa$ and $\overline{\gamma} = \gamma_x + i\gamma_y = \gamma\, e^{2i\theta_{pot}}$ are the usual convergence and shear parameters. We denote $\gamma = \mathcal{D}\tilde{\gamma}$ and $\kappa = \mathcal{D}\tilde{\kappa}$ to separate the distance and mass effects. $\theta_{pot}$ is the direction of the shear (independent of the redshift of the source) and is defined as:

$$\tan(2\theta_{pot}) = \frac{2\partial_{xy}\phi}{\partial_{xx}\phi - \partial_{yy}\phi} \ . \tag{22}$$

Equivalently, for the potential we can define the parameters $\overline{g}_{pot}$, $\overline{\tau}_{pot}$ and $\delta_{pot}$:

$$\overline{g}_{pot} = \frac{\overline{\gamma}}{1 - \kappa} \ , \quad \tau_{pot} = \frac{2\overline{g}_{pot}}{1 - \overline{g}_{pot}\overline{g}_{pot}^*} \ , \quad \delta = 1 + \overline{g}_{pot}\overline{\tau}_{pot}^* \ . \tag{23}$$

For a circular source using this notation, the lens transformation gives: $g_I = g_{pot}$, $\tau_I = \tau_{pot}$, etc.

The determinant of eq. 20 gives the lensing transformation of the object's size:

$$s_S = |\det a^{-1}|\, s_I \ . \tag{24}$$

Dividing eq. 20 by eq. 24 we have the lens equation for the deformation matrix:

$$D_S = \frac{1}{|\det a^{-1}|}\, a^{-1}\, D_I\, {}^t a^{-1} \ . \tag{25}$$

¿From eq. 25, the lens equation for the complex deformation $\overline{\tau}_S$ is also derived:

$$\mathrm{sgn}(\det a^{-1})\overline{\tau}_S = \overline{\tau}_I - \overline{\tau}_{pot}\left(\delta_I - \tau_I\Re(\overline{g}_I\overline{g}_{pot}^*)\right) \ . \tag{26}$$

The inverse equation is found by exchanging the subscripts $I$ and $S$ and the signs of $\overline{\tau}_{pot}$ and $\overline{g}_{pot}^*$. This gives:

$$\mathrm{sgn}(\det a^{-1})\overline{\tau}_I = \overline{\tau}_S + \overline{\tau}_{pot}\left(\delta_S + \tau_S\Re(\overline{g}_S\overline{g}_{pot}^*)\right) \ . \tag{27}$$

A vectorial representation of eq. 27 is shown in Fig. 6. *The complex deformation of the image is just the vector sum of the intrinsic source shape and the induced deformation from the potential,* corrected in the strong lensing regime by a factor $\delta_S + \tau_S\Re(\overline{g}_S\overline{g}_{pot}^*)$. In the weak shear regime ($\det a^{-1} > 0$), the correction tends to unity and eq. 27 becomes:

$$\overline{\tau}_I = \overline{\tau}_S + \overline{\tau}_{pot} \ . \tag{28}$$

Using the local shear axes eq. 26 reads:

$$\mathrm{sgn}(\det a^{-1})\tau_{Sx} = \delta_{pot}\tau_{Ix} - \tau_{pot}\delta_I = \delta_{pot}\delta_I(\varepsilon_I - \varepsilon_{pot}) \ , \tag{29}$$

$$\text{sgn}(\det a^{-1})\tau_{Sy} = \tau_{Iy} \ . \tag{30}$$

Note that $|\tau_y|$ is a conserved quantity under the lens transformation.

## 4.2. Distribution in ellipticity and redshift

The source ellipticity distribution can be estimated from deep HST images of fields outside rich clusters. A large sample of suitable fields are available in the Medium Deep Survey archive (Griffiths et al. 1994). Analysis of these (Ebbels et al. 1995) reveals that the observed distribution of image shapes for brighter galaxies is well fitted by the functional form:

$$p(\tau_{Sx}, \tau_{Sy}) = \frac{1}{2\pi\sigma_\tau^2} \exp\left(-\frac{\tau_{Sx}^2 + \tau_{Sy}^2}{2\sigma_\tau^2}\right) \ , \tag{31}$$

This distribution has a maximum at $(\tau_x, \tau_y) = (0, 0)$ and is also radially symmetric (because of the random orientations of unlensed field galaxies). We stress however that the form of this distribution does depend strongly upon the size of the galaxies and their magnitudes (Ebbels et al. 1995).

Since $|\tau_y|$ is conserved by lensing, *in the frame of the local shear*, we have the conditional probabilities:

$$p(\tau_{Sx}, \tau_{Sy}) = p(\tau_{Sx}, \tau_{Iy}) = p(\delta_{pot}(z_S)\tau_{Ix} - \tau_{pot}(z_S)\delta_I, \tau_{Iy}) = p(z|I, mass)p_{\tau_{Sy}}(\tau_{Sy}) \ . \tag{32}$$

In other words, the conditional redshift probability (given the image shape *and* the mass model) is simply the source shape probability divided by that of $\tau_{Sy}$. From eq. 31:

$$p(z|I, mass) = \frac{p(\tau_{Sx}, \tau_{Sy})}{p_{\tau_{Sy}}(\tau_{Sy})} = \frac{1}{\sqrt{2\pi}\sigma_\tau^2} \exp\left(-\frac{(\delta_{pot}(z_S)\tau_{Ix} - \tau_{pot}(z_S)\delta_I)^2}{2\sigma_\tau^2}\right) \ . \tag{33}$$

which reproduces the intuitive prescription of Kneib et al. (1994a) that the maximum of the redshift probability function for a given image corresponds to the minimum deformation of the source.

When the image is outside the critical line, $\tau_{pot}$ is an increasing function of redshift – with positive $\tau_{Ix}$ if the orientation is within 45deg of the shear direction and negative otherwise. If $\tau_{Ix}$ is positive but not too large, $p(z|I, mass)$ has a maximum for $\varepsilon_{Ix} = \varepsilon_{pot}(z_S)$ and the most probable redshift is finite. However, when $\tau_{Ix}$ is too large, $p(z|I, mass)$ is an increasing function of redshift leading to a most probable redshift, $z = \infty$. If $\tau_{Ix}$ is negative then $p(z|I, mass)$ is a decreasing function of redshift, yielding a most probable redshift, $z = z_{lens}$ (see Fig. 7). In the latter two cases no "sensible" estimate of the redshift of the galaxy can be derived.

### 4.3. Uncertainties in the redshift determination

We now discuss the uncertainties that arise when determining faint galaxy redshifts with a gravitational telescope. There are three sources of error: those arising from image shape (errors in the deformation $\overline{\tau}_I$), the lens mass model (errors in $\kappa$, $\gamma$ and $\theta_{pot}$) and statistical errors introduced by the contamination of the arclet sample by foreground or cluster galaxies. The first two errors are concerned with individual arclets, while the third affects the properties of the sample as a whole.

#### 4.3.1. Individual errors

We begin by considering the relative error in $z$. Differentiating $g_{pot} = \mathcal{D}\tilde{\gamma}/(1 - \mathcal{D}\tilde{\kappa})$, we have:

$$\frac{(z - z_L)\mathcal{D}'}{\mathcal{D}} \frac{dz}{z - z_L} = \frac{d\mathcal{D}}{\mathcal{D}} = (1 - \mathcal{D}\tilde{\kappa})\frac{dg_{pot}}{g_{pot}} - \frac{d\tilde{\gamma}}{\tilde{\gamma}} - \mathcal{D}d\tilde{\kappa} . \tag{34}$$

The term $\mathcal{D}/(z - z_L)\mathcal{D}'$ is almost proportional to $(z - z_L)$, indicating that the accuracy of the lensing-inferred redshifts is lower at large redshift (see Fig. 8).

Moreover, for the maximum of redshift probability function:

$$\frac{dg_{pot}}{g_{pot}} = \delta_{pot}\frac{d\varepsilon_{pot}}{\varepsilon_{pot}} = \delta_{pot}\frac{d\varepsilon_{Ix}}{\varepsilon_{Ix}} = \delta_{pot}\left(\frac{d\varepsilon_I}{\varepsilon_I} - 2\tan(2\theta_I)d\theta_I\right) , \tag{35}$$

and thus the total error in the estimate of the most probable redshift is:

$$\frac{d\mathcal{D}}{\mathcal{D}} = (1 - \mathcal{D}\tilde{\kappa})\delta_{pot}(\frac{d\varepsilon_I}{\varepsilon_I} - 2\tan(2\theta_I)d\theta_I) - \frac{d\tilde{\gamma}}{\tilde{\gamma}} - \mathcal{D}d\tilde{\kappa} . \tag{36}$$

We now turn to the likely uncertainties in measured image ellipticities $\varepsilon_I$ and position angles $\theta_I$. The former has the following effect:

$$\frac{d\mathcal{D}}{\mathcal{D}} = (1 - \mathcal{D}\tilde{\kappa})\delta_{pot}\frac{d\varepsilon_I}{\varepsilon_I} \; . \tag{37}$$

An overestimate of $\varepsilon_I$ leads (in the sub-critical part of the lens) to an over-estimated redshift. If the image is close to a critical line then a small error in $\varepsilon_I$ produces a large error in $z$ since $\delta_{pot}$ diverges. However for the bulk of the arclets this is not a problem as we are not in the multiple image regions and $(1 - \mathcal{D}\tilde{\kappa})\delta_{pot}$ is in general less than 1 (as $\delta_{pot} \sim 1$ and $(1 - \mathcal{D}\tilde{\kappa}) < 1$). For faint images there is a tendency to under-estimate the image ellipticity at large ellipticity and over-estimate the ellipticity at small ellipticity, although these effects can be statistically corrected. However for small ellipticities and compact objects it is difficult to determine the true image ellipticity.

Errors in the measured orientation $\theta_I$:

$$\frac{d\mathcal{D}}{\mathcal{D}} = -(1 - \mathcal{D}\tilde{\kappa})\delta_{pot}2\tan(2\theta_I)d\theta_I \; . \tag{38}$$

have the same dependence on $\delta_{pot}$ as the ellipticity. However, as the orientation is usually the best measured characteristic of an image and because the error is symmetrically distributed, the bias is less serious than for the ellipticities. Nevertheless, when $\theta_I \sim \pi/2$ ($\varepsilon_{Ix} \sim 0$) the errors can become very large.

Finally, the errors in the cluster mass model, in $\tilde{\kappa}$:

$$\frac{d\mathcal{D}}{\mathcal{D}} = -\mathcal{D}d\tilde{\kappa} = -\frac{d\Sigma}{\Sigma_{crit}} \; , \tag{39}$$

and the error in $\tilde{\gamma}$:

$$\frac{d\mathcal{D}}{\mathcal{D}} = -\frac{d\tilde{\gamma}}{\tilde{\gamma}} \; . \tag{40}$$

demonstrate that an over-estimated local mass and shear lead to an under-estimated redshift, with the dependence between the two being reasonably well-behaved. However, adding galaxy-size components in the lens-model can dramatically change the intensity of the shear near a critical area of the lens, therefore it is important to take these components into account.

An advantage of the Hubble Space Telescope is the stability of the high resolution imaging, this minimises problems which plague ground-based studies of faint object shapes (seeing, tracking, field astigmatism, and their time variability). While the HST capabilities are an order of magnitude better than the ground-based facilities, the limiting factors for accurate measurement of the shapes of faint and compact galaxies now become photon noise and pixel-sampling effects.

A key point in determining reliable redshifts is the *absolute* calibration of the mass model. This is best addressed using a number of spectroscopically-confirmed lensed features in the cluster, while the morphology of the mass can be best determined using the geometry of any multiply-imaged sources present (Mellier et al. 1993, Kneib et al. 1993, Smail et al. 1995c). For spectroscopic arcs, Abell 2218 is one of the best clusters for our purposes since Pello et al. (1992) have secured accurate redshifts for two of the giant arcs in the cluster core. Similarly, the presence of at least 7 multiply-imaged sources in Abell 2218, identified using HST, means that we can strongly constrain not only the absolute mass in the cluster core, but also the detailed form of its distribution. We can thus expect that remaining uncertainties in the mass distribution will predominantly arise from unresolved granularity on scales $\leq 75$ kpc not attached to any galaxy. The fact that we can make such a statement attests to the detailed view of the cluster mass provided by lensing.

### 4.3.2. Sample selection contaminations

The final uncertainty we must consider arises from contamination of the arclet catalog by foreground galaxies and, in particular, cluster members. Indeed, considering only the number of galaxies detected within the WFPC field and comparing this to deep field counts (Smail et al. 1995d), ~30 galaxies per magnitude are cluster members down to R~24.5 (Fig. 2). This contamination is stronger in the center of the cluster than in the outer parts as the surface density of galaxies within a cluster falls faster than $1/r$.

If the contaminating galaxies are randomly orientated, the shear will be reduced below the true value and hence we obtain an artificially reduction of the mean redshift of the background population. In the absence of a reliable distance separation on the basis of arclet colors, we have developed a statistical method to estimate the unlensed contamination.

Lensing displaces the ellipticity distribution of faint sources from that observed for blank fields (or unlensed sources) which should be centred on the null ellipticity; this is illustrated schematically in Fig. 9. As discussed in §4.2, redshifts can only be estimated for images with orientations within 45deg of the predicted shear direction (i.e. arclets with $\tau_{Ix} > 0$). Images with $\tau_{Ix} < 0$ must either be cluster members, foreground galaxies or, conceivably, lensed background objects which are insufficiently deformed to move them into the $\tau_{Ix} > 0$ region. The number of images with $\tau_{Ix} < 0$ therefore provides an upper limit on contamination by unlensed galaxies and an improved estimate can be determined by considering those lying within the bulk of the ellipticity distribution ($\tau_I < \sigma_\tau$ and $\tau_{Ix} < 0$). By applying a $\pi/2$ rotation prior to inversion, we can also obtain an estimate of the contamination as a function of redshift and directly subtract this spurious $N(z)$ from that derived for the total distribution. Although this only provides a statistical correction for contamination, it gives a good indication of the stability of the derived $N(z)$.

## 5. Determining the Field Redshift Distribution to $R \simeq 25.5$

We now use the photometric catalog of faint arclet candidates discussed in §2, together with the lensing inversion method presented in §4, to derive the likely redshift distribution $N(z)$ of faint background galaxies viewed through the center of Abell 2218.

In order to quantify the errors in our determination of $N(z)$ we must first estimate uncertainties in the shape measurements of individual arclets as a function of their size and apparent magnitude. This is important in determining the useful limit of our HST image for accurate inversion. To do this we simulated $\sim 300\,000$ images of different known sizes and ellipticities and then reproduced the detection characteristics applicable to our HST image. By comparing the actual galaxy catalog to its simulated equivalent, the dispersion in the realised shape and orientation can be examined as a function of apparent magnitude (Fig. 10). The formalism of §4 then gives, for each image, the likely redshift error arising from these observational uncertainties.

The simulations are very helpful in revealing two important limitations that apply in deriving redshift distributions from lensing data:

Firstly, we find the redshift error does not track the measured apparent magnitude very well for a realistic distribution of image properties, but depends more closely on the intrinsic shape and signal/noise of each image. Clearly the most interesting region for consideration is that which lies beyond the current spectroscopic limit, viz. $R > 23$. In the context of our relatively short HST exposure of Abell 2218, Fig. 10d shows that the uncertainty in the measurement of ellipticities increases significant beyond $R = 25$ (but strongly depend on the size and the ellipticity of the objects) and thus inversion becomes highly uncertain. Although we can correct for this effect *statistically* (see §4.3.2, Fig. 11), clearly the uncertainty in this correction could swamp the signal from those sources for which reliable inversion is possible.

Secondly, even with adequate signal/noise for all images, $\Delta z$ depends on $z$ itself (Fig. 8). A single cluster lens can thus only usefully constrain the number of sources lying in a specific redshift range ($0.5 < z < 1.5$ for Abell 2218) although some information is available on the overall $N(z)$ as well.

The first limitation is more serious as it emphasises that those samples for which lensing-induced redshift distributions can be reliably determined are unlikely to be strictly magnitude-limited as has been the case traditionally for ground-based spectroscopic surveys. Notwithstanding the contamination from sources which are not amenable to inversion, a magnitude-limited arclet sample would never produce a magnitude-limited source sample because of the variable magnifications. However, the fact that little can be said about a subset of faint sources is a more serious difficulty when comparing with current model predictions which are largely based on integrated magnitudes. Either it must be assumed that the compact sources are a representative subset of those for which inversion is practical, or evolutionary models must take into account the effect of an areal threshold rather than an integrated magnitude. Conceivably with much longer integrations, the signal/noise of each faint image will improve sufficiently to reduce the uncertainties.

As the source surface brightness and $k$-corrections depend strongly on redshift, the visibility of a faint source is also a complex function of redshift and type. Although this is true of any isophotally-selected faint galaxy sample and is not further distorted by the lensing process, as the arclet population presumably probes to much higher redshift than the brighter spectroscopic samples, the uncertainties in allowing for visibility losses are presumably much greater. A specific problem, raised originally by Smail et al. (1991), is the possibility that only dense star-forming regions have sufficient ultraviolet flux and high enough surface brightness to produce arclets visible with HST.

We now illustrate the above effects in the context of the actual Abell 2218 catalogue. To $R \simeq 25$, we have 235 candidate arclets and for each of these, our procedure delivers a likelihood distribution for the true, unlensed apparent magnitude $R_{source}$ and the redshift $z$. We can apply the contamination correction discussed in §4.3.2 to determine the mean redshift of those sources with $z > z_{cl}$ and this can be compared with various predictions. This method is illustrated on the $R_{source}-z$ plane in Figure 11 together with the no evolution prediction for an $R$-limited sample following the procedure described by Ellis (1995). The latter prediction is based on type-dependent $b_J$ luminosity functions and morphological proportions observed for the local field population (Loveday et al. 1992) transformed to the $R$-band using Hubble-sequence colours with $k$-corrections taken from King & Ellis (1985). The results are also summarised in Table 4.

For $R < 22$ there are too few arclets in our catalogue for meaningful results, but for $22 < R < 25$ the results indicate a gradual increase in mean redshift with apparent magnitude (Fig. 11 dashed-line). The mean redshift of the arclet population is reasonably close to the no evolution expectations to $R \simeq 24$. However, upon examination of the individual redshifts, there appears to be an excess of low redshift arclets whose proportion is independent of magnitude and whose origin could explain the trend towards low mean redshifts for faint arclets found earlier by Smail et al. (1995b) and Kneib et al. (1994a). It is now clear, following the discussion above, that this effect arises because a fraction of the images have insufficient shear to be correctly inverted and the residual uncertainties in the correction illustrated in Fig. 11 can affect the results at the level where interesting scientific conclusions are required. We can quantify this effect by restricting the technique to those images whose isophotal areas exceed 50 pixels and ellipticity is larger than 0.2. As Fig. 11 shows (dotted-dashed line), this leads to an increase in the mean redshift at all magnitudes and the large majority of the 'low-z' points disappear.

Out of a total sample of 42 well-defined arclets to $R \simeq 25.5$, only two are beyond $z \simeq 2$ and the mean redshift at $R \simeq 25$ is $\simeq 1$ i.e. only slightly above the no evolution prediction. Thus, unless the smaller arclets represent an entirely different population of sources or the intrinsic size of a source is a strong function of redshift beyond $\simeq 1$, the absence of a large number of very distant luminous sources to $R \simeq 25$ is a secure result.

It is important to recognise that, for a lensing cluster at $z{=}0.175$, Fig. 8b shows the mean error in inverted redshift is high even for a well-defined arclet. Typically, for the arclets amenable to individual inversion, $\Delta\epsilon/\epsilon \simeq 0.05$-$0.1$ and thus, from Fig. 9b, a source at $z \simeq 2$ could be placed anywhere from $1{<}z{<}3$. A large sample size obviates the need for more precise inversion but we also note that the redshift distribution would be verified via inversion through a well-constrained cluster at higher redshift. An arclet sample viewed through a cluster at $z \simeq 0.3$ would reduce the implied redshift error for a source at $z \simeq 2$ by a factor of 2.

It may also be possible to verify the brighter arclets spectroscopically and this will clearly lead to further improvements. Such confirmation can be made more effective by selecting the bluest cases with predicted redshifts $z \leq 2$. where strong emission lines would lie in the range of optical and near-infrared spectrographs. A subset of well-distributed arclets would represent a valid test of our inversion since this depends on geometrical quantities and the cluster mass model, both of which are independent of the photometric properties of the background sources. As an encouragement for interested workers, we therefore list in Table 2 the inverted redshifts for each of the major arcs and multiple images discussed in §2.3 and those remaining arcs for which reliable inversion is possible plotted on Fig. 11.

Finally, we turn to the effect on the inversion of our improved HST-based mass model compared to the earlier ground-based equivalent of Kneib et al. (1995). In Table 4 we show the mean inverted redshift after statistical correction for contamination for the ground-based ('old') and HST-based ('new') mass models. In both cases, the differences are very minor and illustrate that the principal uncertainty in inversion through Abell 2218 is no longer the global mass model for the cluster.

Although inversion through well-constrained lenses is an extremely promising prospect, this pilot study has shown that the major observational limitation is the signal/noise of the required shape parameters for the faint sources. This demonstrates the importance of securing deeper HST exposures. A second revelation is the importance of developing a new approach in the construction of model predictions. It seems unlikely that such faint sources can easily be constructed into apparent magnitude-limited samples and thus much work is needed to produce surface brightness limited predictions. Notwithstanding these difficulties, Abell 2218 remains an exceptionally promising cosmic lens and the opportunities for verifying or otherwise the predicted redshifts for the brighter arclets are excellent. Such data will improve the mass model and lead to even tighter constraints on the redshifts of sources beyond reach of ground-based spectrographs.

## 6. Conclusions

1. This paper is the first attempt to constrain the redshift distribution of very faint galaxies by using the HST to study images of sources which have been lensed by a massive foreground cluster. Using multiple images and newly-discovered lensing features in the rich cluster Abell 2218, we have constructed a precise mass distribution which is more tightly constrained that for any other cluster. With HST high-resolution data we have now sufficient information to constrain not only the mass profile of the cluster but also to give limits on the masses of individual cluster galaxies.

2. Using this mass model and a new formalism we develop based on the observed image parameters of faint sources, we demonstrate how it is possible to deduce the redshift distribution of very faint galaxies viewed through the cluster as well as to account statistically for contamination by unlensed sources and in understanding the various uncertainties. First results are presented for a large sample of arclets to $R \simeq$25-25.5, $\sim$3 magnitudes fainter than results from ground-based spectroscopy.

3. Several limitations arise in the interpretation of our inverted redshifts when comparison is made with evolutionary models. These limitations may help to explain earlier results which have tended to yield arclet redshifts somewhat less than extrapolation of ground-based spectroscopic data would imply. Even with the high resolution of the HST, it is difficult in short integrations to measure faint galaxy shapes adequately to invert a magnitude-limited sample. It therefore appears more practical to invert area-limited samples and we demonstrate that more reliable results are obtained using such a subset.

4. We demonstrate that, notwithstanding the uncertainties and sample selection criteria we have adopted, the absence of a large number of very distant sources ($z > 2$) in our inverted redshift distributions is a robust result. At $R \simeq$25, the mean redshift for samples corrected for contamination or those based on individual arclets of high signal/noise is only $\simeq$0.8-1.2.

5. The brighter arclets, whose redshifts are estimated via our technique, are amenable to direct spectroscopic examination. Such confirmation can be made more effective by selecting the bluest cases with predicted redshifts $z \leq 2$. where strong emission lines would lie in the range of optical and near-infrared spectrographs. Note that confirmation of a carefully-selected subset of well-distributed arclets would still represent a valid test of our inversion technique. The lensing inversion depends only on geometrical quantities and the cluster mass model, both of which are independent of the photometric properties of the background sources.

We thank Bob Williams and Ray Lucas and other STScI staff for their assistance with the rapid processing of this data. We acknowledge enthusiastic support from a number of colleagues including Roger Blandford, Bernard Fort, Yannick Mellier, Jerry Ostriker, Roser Pelló, Peter Schneider and Martin Rees. This paper is based on observations with the NASA/ESA *Hubble Space Telescope*, obtained at the Space Telescope Science Institute, which is operated by the Association of Universities for Research in Astronomy Inc. JPK gratefully acknowledges support from an EC Fellowship and IRS from NATO and Carnegie Fellowships. WJC acknowledges support from the Australian Department of Industry, Science and Technology, the Australian Research Council and Sun Microsystems.

| Multiple images | R (F702W) | B-r* | $\mu_{mean}$ | z |
|---|---|---|---|---|
| 384 / 468 | 21.2 / 22.6 | 1.0 | 23.35 / 23.50 | $2.8^{+0.5}_{-0.2}$ |
| 359/328/337/389 | 20.3/22.0/21.9/21.5 | 3.1/3.6/2.6/2.5 | 22.65/22.65/22.70/22.80 | 0.702 |
| 289 | 20.5 | 0.75 | 23.10 | 1.034 |
| 730 | 22.9/22.8/ 23.8 (21.9) | 1.43 | 23.90/23.70/23.45 | 1.1±0.3 |
| H1-2-3 | ∼25.5 | — | ∼24.2 | 1.0±0.3 |
| H4-5 | ∼26.0 | — | ∼24.5 | 1.6±0.3 |
| 444 / H6 | 22.7 / 23.6 | 0.35 / — | 23.85/23.70 | 1.1±0.1 |

* from Leborgne et al. (1992) when available, typical errors for red objects can be as high as 0.5 mag (Kneib et al. 1995).

Table 1: List of confirmed and candidate multiple images from our HST study, along with their ground-based colors and measured or predicted redshifts.

| # id | $R_{cor}$ | $\mu_R$ | $z_{photo}$ | z- | $z_{opt}$ | z+ | Comments |
|------|-----------|---------|-------------|-----|-----------|-----|----------|
| 190* | 22.1 | 23.4 | 0.1-0.7 | — | — | — | disk galaxy |
| 231* | 22.2 | 23.1 | 0.2-0.7 | 0.3 | 0.4 | 0.6 | compact + disk galaxies |
| 238 | 23.9 | 23.5 | 2.0-3.0 | 0.8 | 1.2 | 1.6 | disk galaxy |
| 254 | 24.1 | 23.9 | 0.2-0.7 | 0.4 | 0.6 | 0.7 | |
| 289 | 22.3 | 23.1 | — | — | 1.034 | — | |
| 300* | 24.5 | 23.6 | 0.8-2.5 | — | — | — | 2 compact galaxies |
| 309 | 25.2 | 24.0 | 1.4-2.0 | — | — | — | |
| 323* | 21.1 | 22.6 | 1.4-1.6 | 0.2 | 0.4 | 0.6 | |
| 344* | 22.9 | 23.8 | 1.8-3.0 | — | — | — | 2 extended galaxies |
| 359 | 24.9 | 22.7 | — | — | 0.702 | — | |
| 362 | 25.8 | 24.0 | 1.8-3.0 | 0.5 | 1.1 | 2.0 | |
| 365* | 23.7 | 23.8 | 2.4-3.0 | 0.3 | 0.3 | 0.4 | |
| 382 | 25.7 | 24.0 | 1.4-2.6 | 2.5 | 3.0 | 3.5 | |
| 384 | 24.6 | 23.3 | 2.6-3.5 | 2.6 | 2.8 | 3.3 | |
| 444 | 25.4 | 23.5 | 1.8-3.0 | 1.0 | 1.1 | 1.2 | |
| 456 | 24.5 | 23.4 | 0.1-0.7 | 0.4 | 0.6 | 0.8 | |
| 467 | 21.7 | 22.1 | 1.2-2.6 | 0.4 | 0.4 | 0.5 | |
| 468 | 23.6 | 23.5 | 1.8-3.0 | 2.6 | 2.8 | 3.3 | counter-image of 384 |
| 730 | 25.1 | 23.7 | 2.9-3.3 | 1.0 | 1.1 | 1.2 | |
| 731 | 26.0 | 23.9 | 2.3-3.1 | 0.8 | 1.1 | 1.4 | |
| 273 | 23.8 | 22.2 | — | 0.5 | 0.6 | 0.7 | |
| H1,2,3 | ∼27. | ∼24.2 | — | 0.7 | 1.0 | 1.3 | cusp arc, disk of 273? |
| H4,5 | ∼27.5 | ∼24.5 | — | 1.3 | 1.6 | 1.9 | fold arc |
| 328 | 24.9 | 22.6 | — | — | 0.702 | — | counter-image of 359 |
| 337 | 24.9 | 22.7 | — | — | 0.702 | — | counter-image of 359 |
| 389 | 24.9 | 22.8 | — | — | 0.702 | — | counter-image of 359 |
| 200 | 24.9 | 23.2 | — | 0.8 | 1.0 | 1.3 | |
| 229 | 24.8 | 23.8 | — | 0.8 | 1.0 | 1.2 | |
| 230 | 24.1 | 23.3 | — | 0.3 | 0.3 | 0.4 | |
| 236 | 23.0 | 23.3 | — | 0.4 | 0.4 | 0.5 | |
| 297 | 23.8 | 23.7 | — | 0.5 | 0.6 | 0.7 | |
| 308 | 23.4 | 23.4 | — | 0.5 | 0.6 | 0.8 | |

| Cluster-Size Component | $x_c$ (arcsec) | $y_c$ (arcsec) | a/b | $\theta$ degree | $r_c$ (kpc) | $\sigma$ (km/s) | $r_{cut}$ (kpc) |
|---|---|---|---|---|---|---|---|
| # 391 | 0.0 | 0.0 | 1.37 | -10. | 76. | 1335. | 710. |
| # 244 | -3.0 | -67.0 | 1.26 | 111. | 33. | 495. | 450. |
| # 196 | -22.0 | -69.0 | 1.05 | 56. | 50. | 470. | 270. |
| # 235 | 26.0 | -104.5 | 1.22 | 152. | 30. | 306. | 200. |

| Galaxy-Size Component | $r_c$ (kpc) | $\sigma$ (km/s) | $r_{cut}$ (kpc) | $M_{tot}$ $10^{12}$ $M_\odot$ | $M/L_V$ $M_\odot/L_\odot$ |
|---|---|---|---|---|---|
| for $M_V^*=-23.$ | 1.0 | 245.0 | 30. | 1.3 | 9. |

Table 3: Characteristics of the various components in the improved mass model of Abell 2218. Positions and orientations are defined on the WFPC-2 image (Fig. 1a) with the position angle $\theta$ increasing anti-clockwise from the X-axis.

| R | $<z>_{NE}$ | $N_{arclet}^{total}$ | $N_{arclet}^{corr}$ | $<z>_{old}$ | $<z>_{new}$ | $<z>_{new}^{nocorr}$ |
|---|---|---|---|---|---|---|
| 22-23 | 0.41 | 11 | 8 | 0.49 | 0.49 | 0.43 |
| 23-24 | 0.53 | 28 | 26 | 0.57 | 0.68 | 0.65 |
| 24-25 | 0.66 | 31 | 31 | 0.64 | 0.83 | 0.83 |

Table 4: Inversion results using all arclets: $<z>_{NE}$ represents the mean expected for no luminosity evolution. $N_{arclet}^{total}$ is the total number in each subsample, $N_{arclet}^{corr}$ is that after correction for contamination by unlensed images. $<z>_{old}$ and $<z>_{new}$ represent, respectively, the mean arclet redshift using the ground-based (Kneib et al. 95) and HST-based mass models. For the HST model, $<z>_{new}^{nocorr}$ indicates the mean redshift prior to contamination correction.

---

Fig. 1.— (a) The full field of our F702W WFPC-2 exposure of Abell 2218 ($z = 0.175$). (b) Central portion showing that several multiply-imaged sources, numbered according to the scheme of Le Borgne et al. (1992), are confirmed by virtue of their mirrored morphological features (see text for details).

Fig. 2.— Differential galaxy counts within the Abell 2218 WFC image (4.7 arcmin$^2$). The dashed line defines our estimated completeness limit at the 55% level. The dotted line indicates field counts in R from Smail et al. (1995d). The dotted-dashed line is the cluster galaxies counts estimated by subtracting the field counts from the observed counts.

Fig. 3.— The distribution of $\sim 120$ arclet candidates with $22 < R < 26$ selected from the HST image (thin lines) compared to those in the ground-based analysis of Pelló et al. (1992) (thick lines). The shear field based on the HST sample illustrates the need for further mass components associated with the brighter cluster galaxies.

Fig. 4.— Shear map for the cluster center derived from the orientations and ellipticities of the HST arclets. The most significant mass components are indicated. The new mass model extends that of Kneib et al. (1995) by including major mass components associated with galaxies #196 and #235 (see Table 3) as well as smaller halos around 30 luminous cluster members (see text for details). At the cluster redshift 1 arcsec is equivalent to 3.83 kpc.

Fig. 5.— Contour map for the adopted mass distribution and the shear map implied for a source plane at $z_S = 1$. Countours correspond from the lowest to the highest to a density of 0.5, 1., 1.5, 2., 3., 4., 5., 6., 8. and 10. $10^9$ M$_\odot$ kpc$^{-2}$. At the cluster redshift 1 arcsec is equivalent to 3.83 kpc.

Fig. 6.— Lens deformation diagram (see text for definition of quantities).

Fig. 7.— Redshift probability distribution for $\tau_{Ix} > 0$ (full and dashed line) and $\tau_{Ix} < 0$ (dotted line) – see text for details.

Fig. 8.— (a) $\mathcal{D}$ parameter versus source redshift for cluster lenses at $z = 0.175$ and $z = 0.35$. The solid line is for $\Omega_0 = 1$ and the dashed line for $\Omega_0 = 0.2$. (b), $\mathcal{D}/[(z - z_L)\mathcal{D}']$ versus source redshift illustrating the higher the source redshift, the greater the uncertainty in the inverted redshift. Note that the redshift of a distant source is more accurately derived using a high redshift lens.

Fig. 9.— Ellipticity distribution showing that observed is the sum of the lensed and unlensed galaxies vectors.

Fig. 10.— Simulated errors for image parameters relevant to lensing inversion as function of apparent magnitude: (a) dispersion on the ellipticity $\sigma_\varepsilon$, (b) dispersion on the orientation $\sigma_\theta$, (c) relative error $\sigma_\varepsilon/\varepsilon_{mes}$ and (d) relative error $(\varepsilon_{true} - \varepsilon_{mes})/\varepsilon_{mes}$. Each data point was determined from 100 realisations of the same source., a dot denotes galaxies with small ellipticities ($\epsilon < 0.2$), (*) denotes galaxies whose isophotal area <50 pixels, and (+) denotes galaxies with intrinsic large ellipticities ($\epsilon > 0.2$) and large isophotal area >50 pixels.

Fig. 11.— Mean redshift vs. intrinsic magnitude for various arclet samples. The solid line represents the no-evolution prediction according to assumptions detailed in the text. The dashed line represents the results for all arclet candidates after making a statistical correction for foreground and cluster contamination. The dotted-dashed line is the same sample after excluding images whose isophotal areas are smaller than 50 pixels. Squares represent the individual inverted redshifts of all arclets greater than 50 pixels in area. Solid symbols denote those with $\epsilon > 0.3$ and open symbols those with $\epsilon < 0.3$.

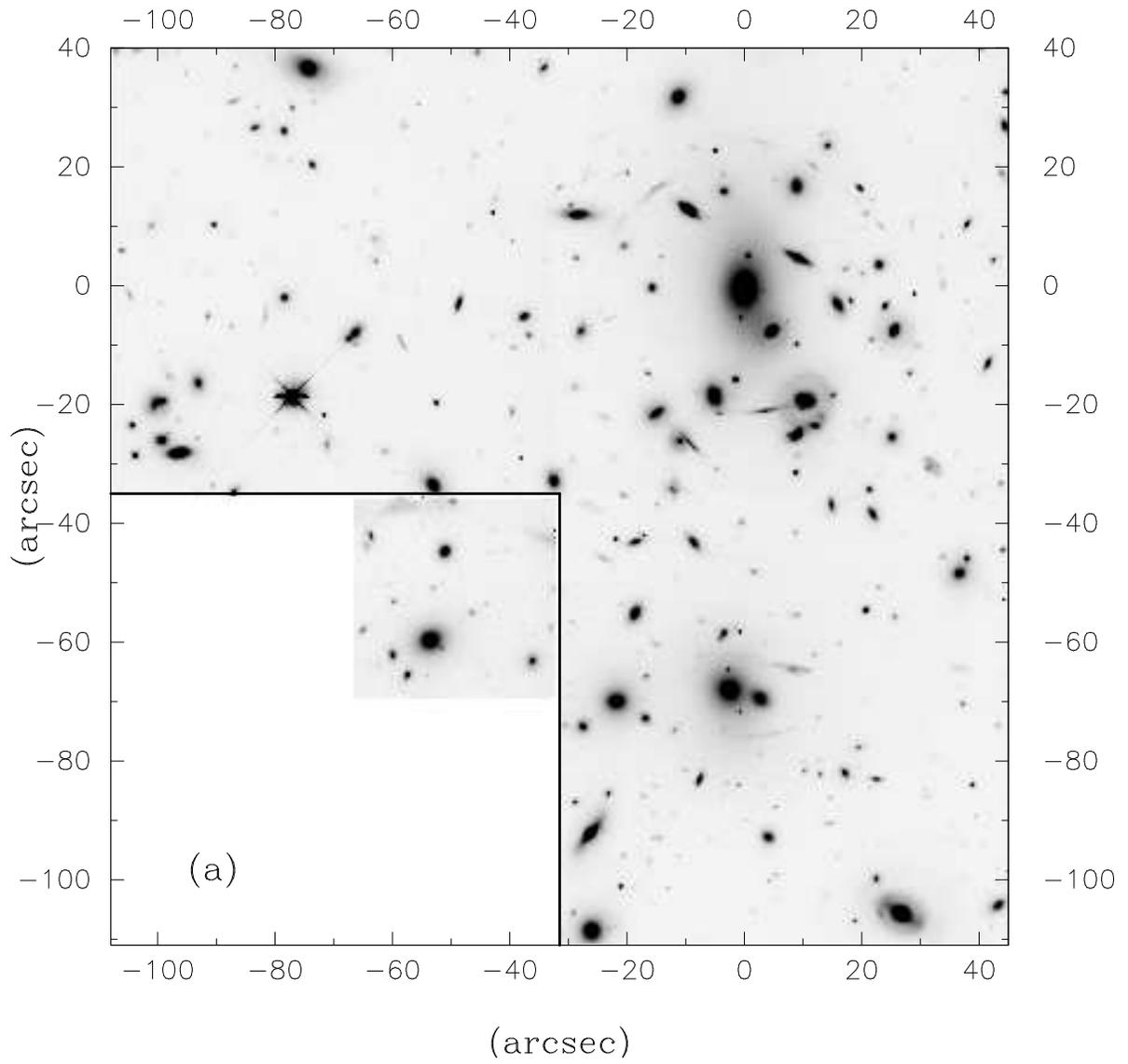

**Figure 1a**

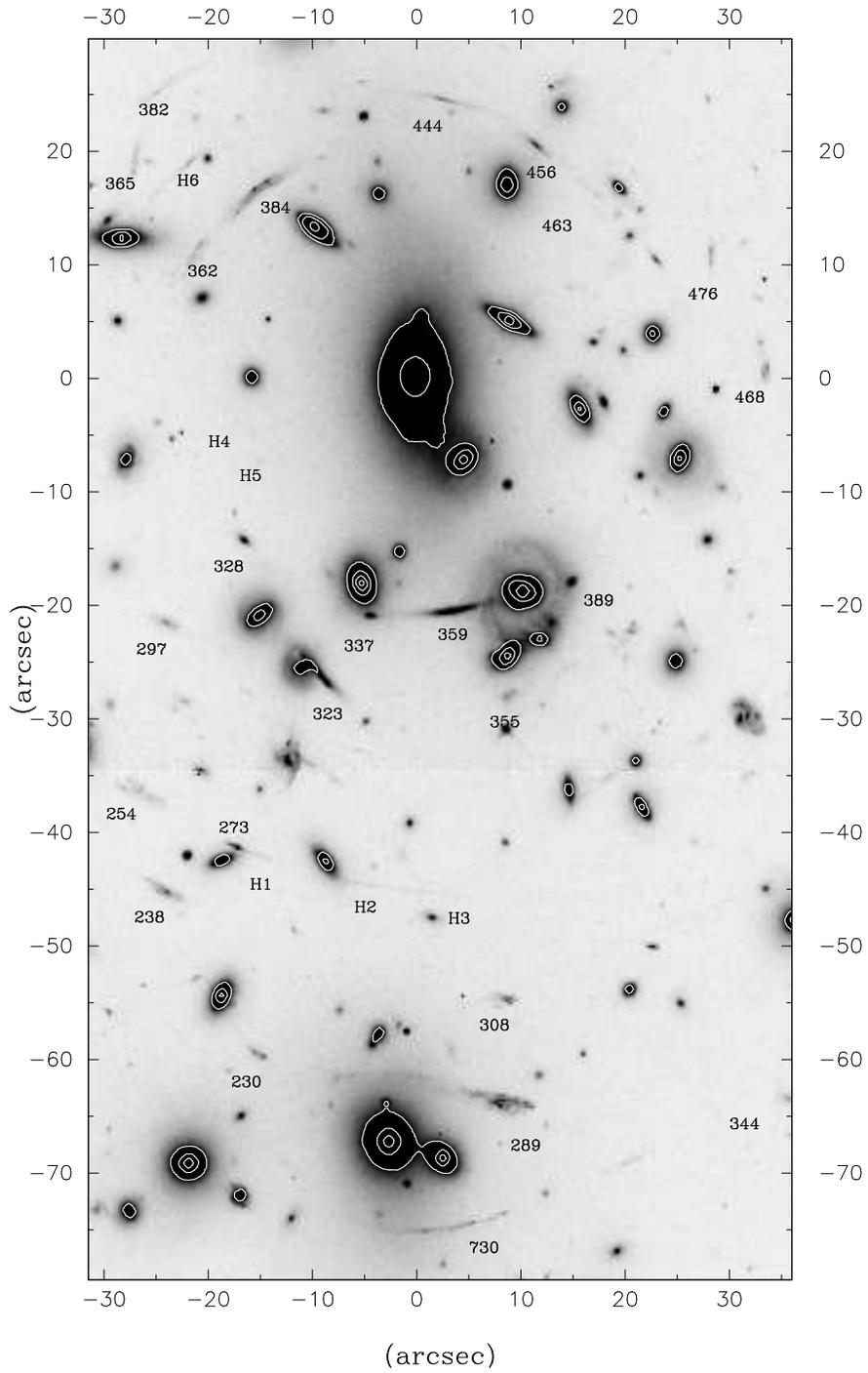

**Figure 1b - Plate**

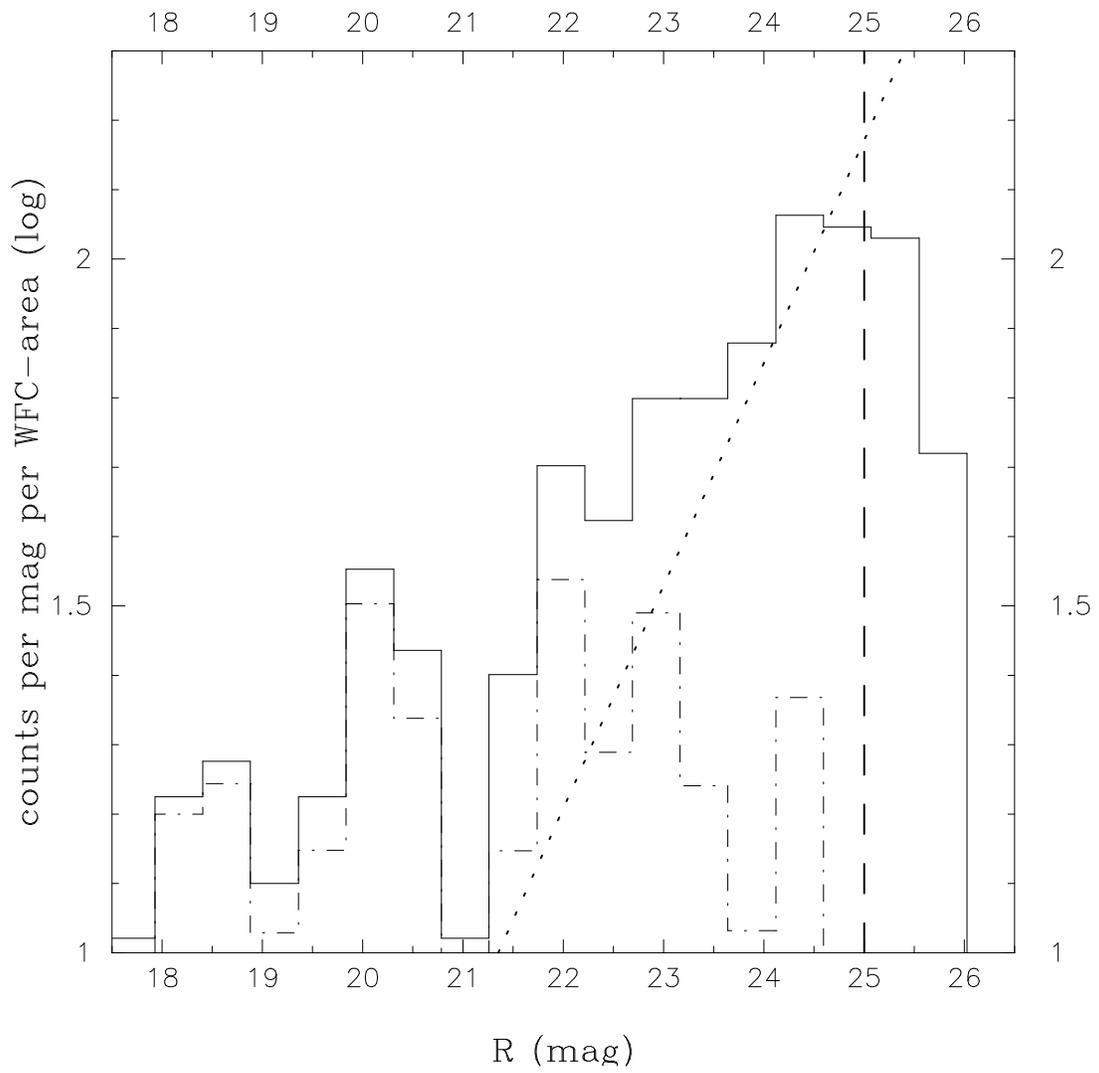

**Figure 2**

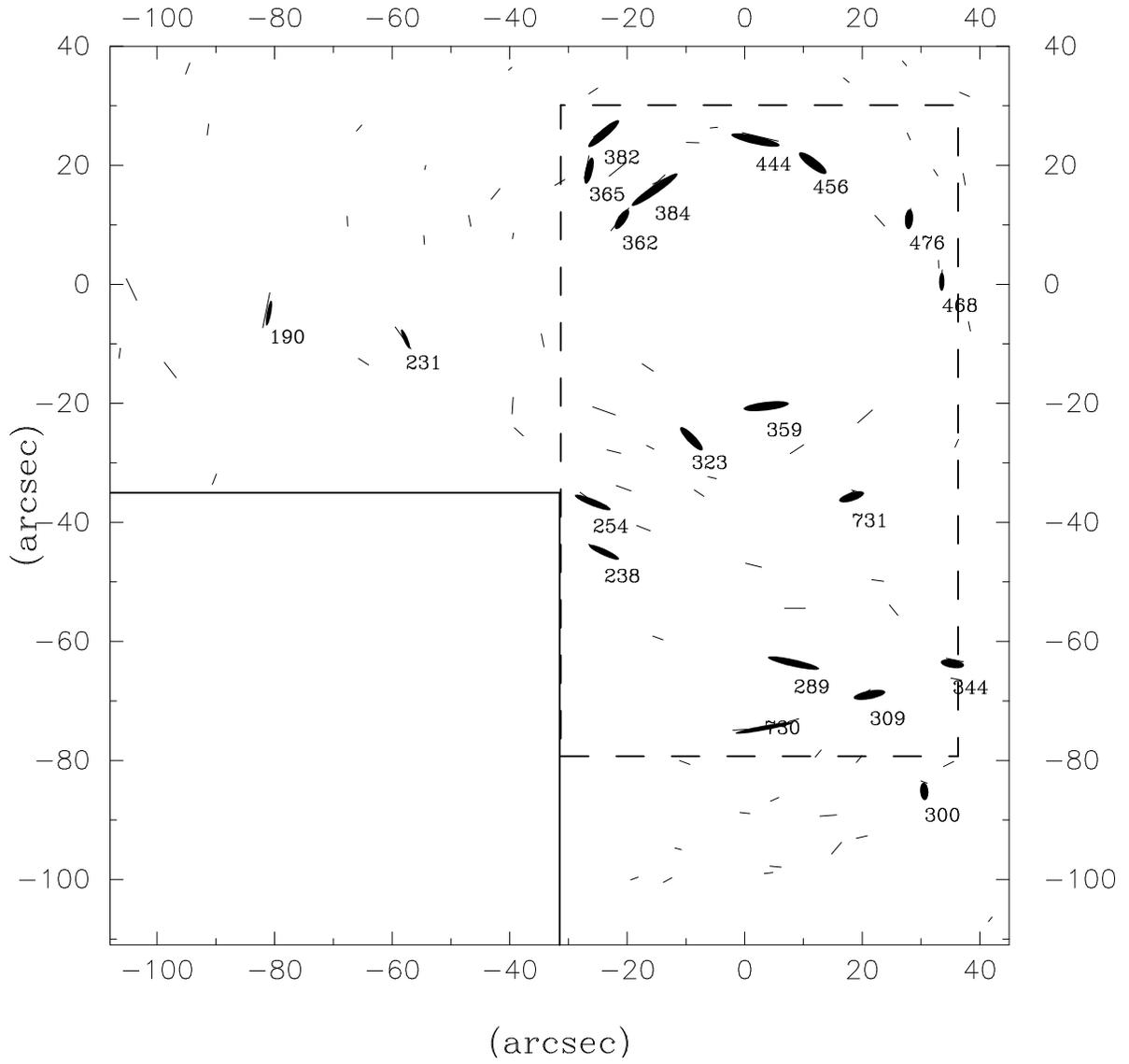

**Figure 3**

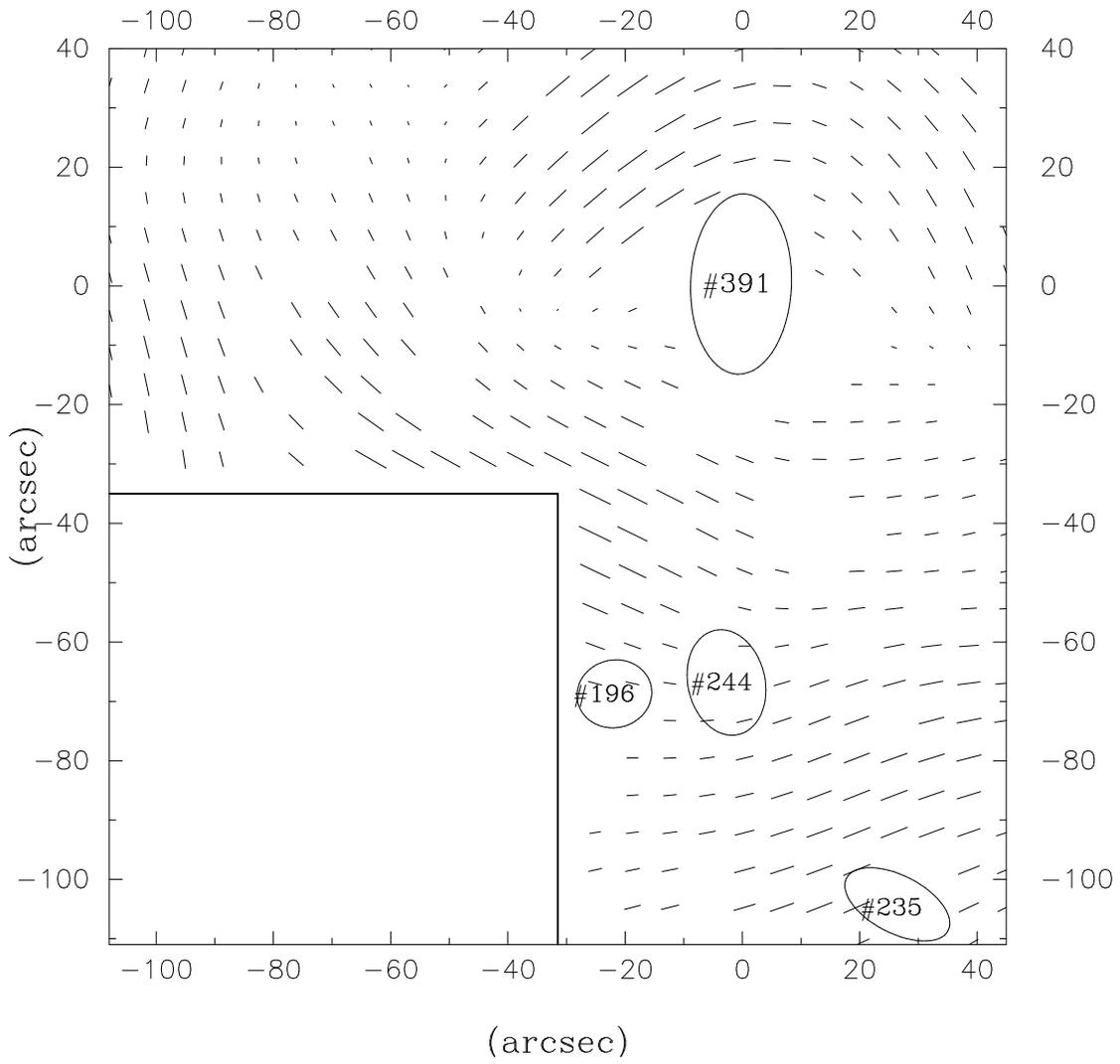

Figure 4

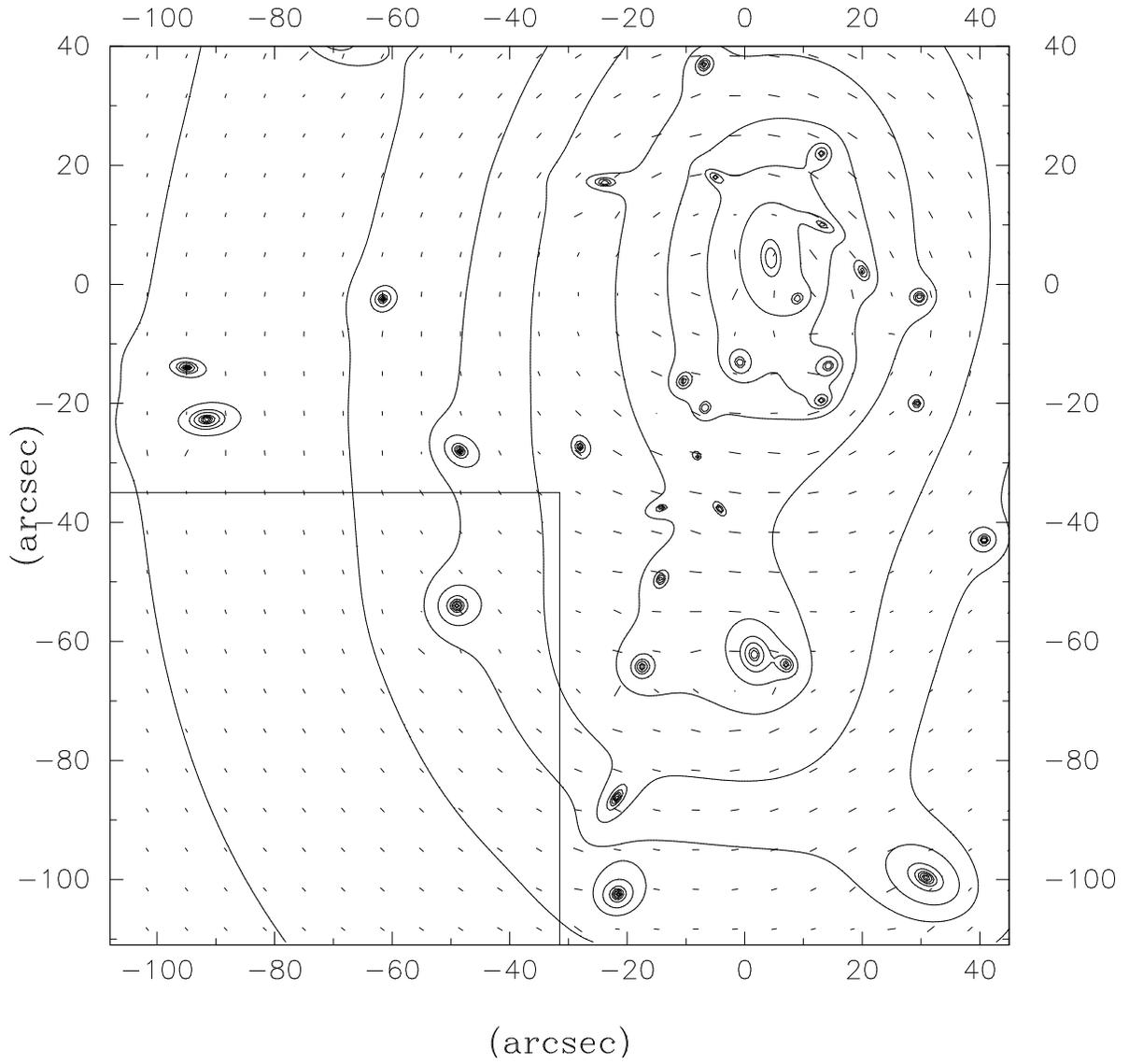

**Figure 5**

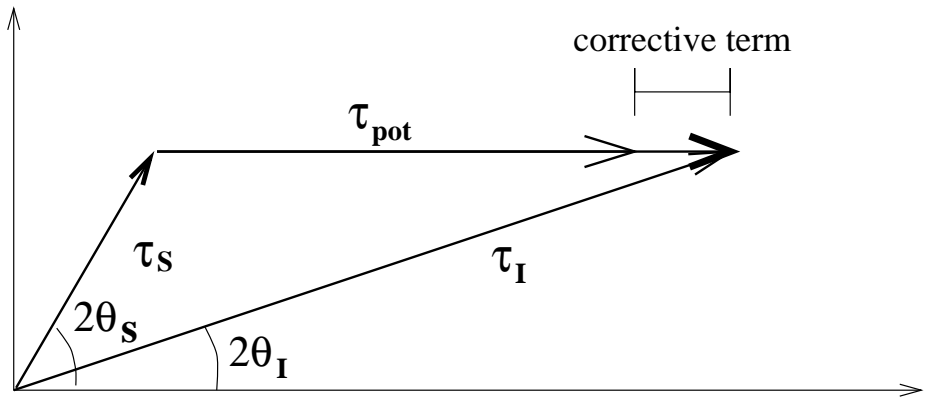

Figure 6

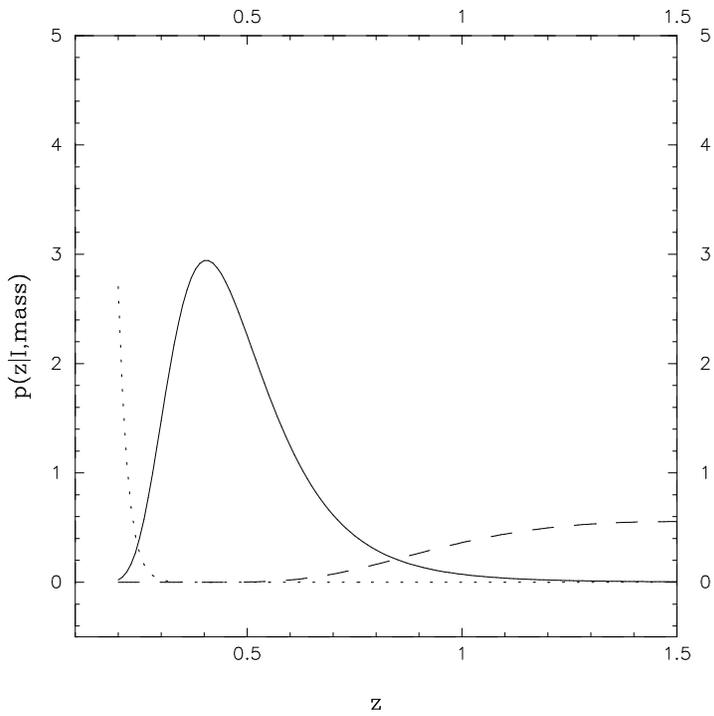

**Figure 7**

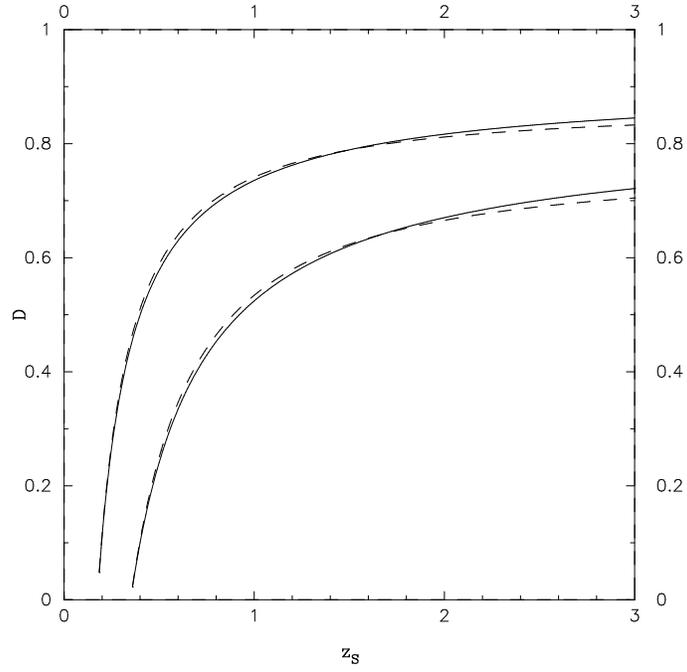

**Figure 8a**

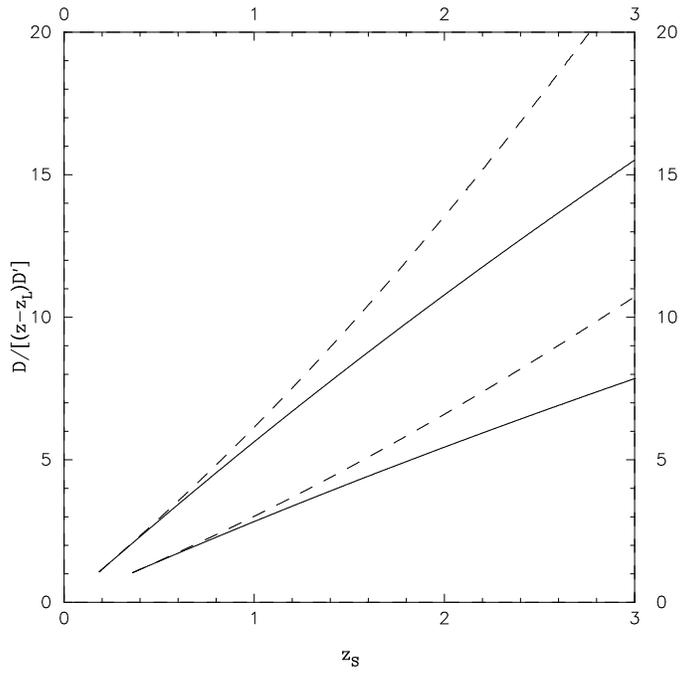

**Figure 8b**

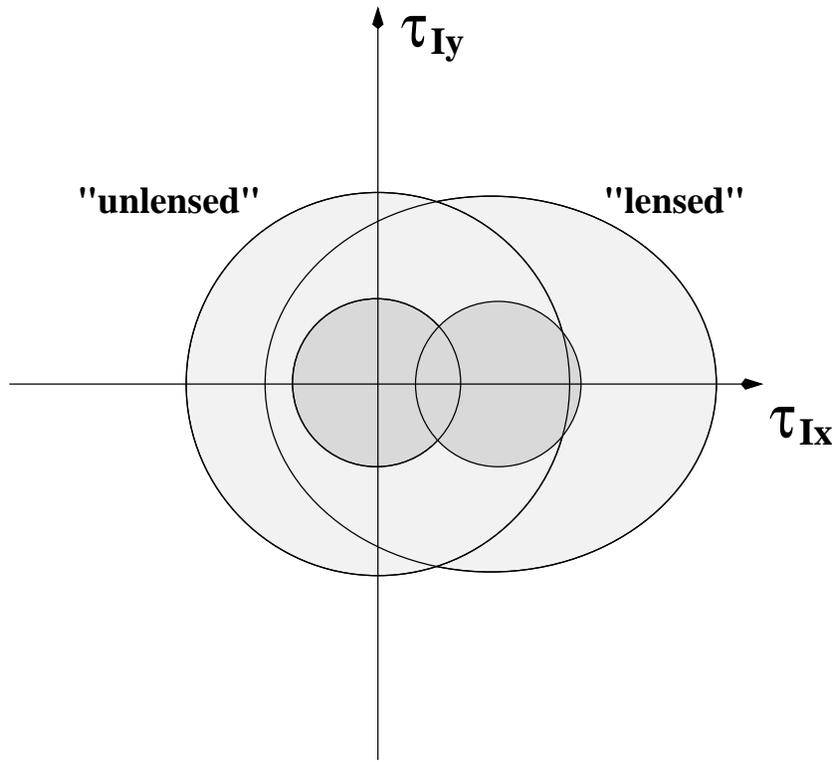

Figure 9

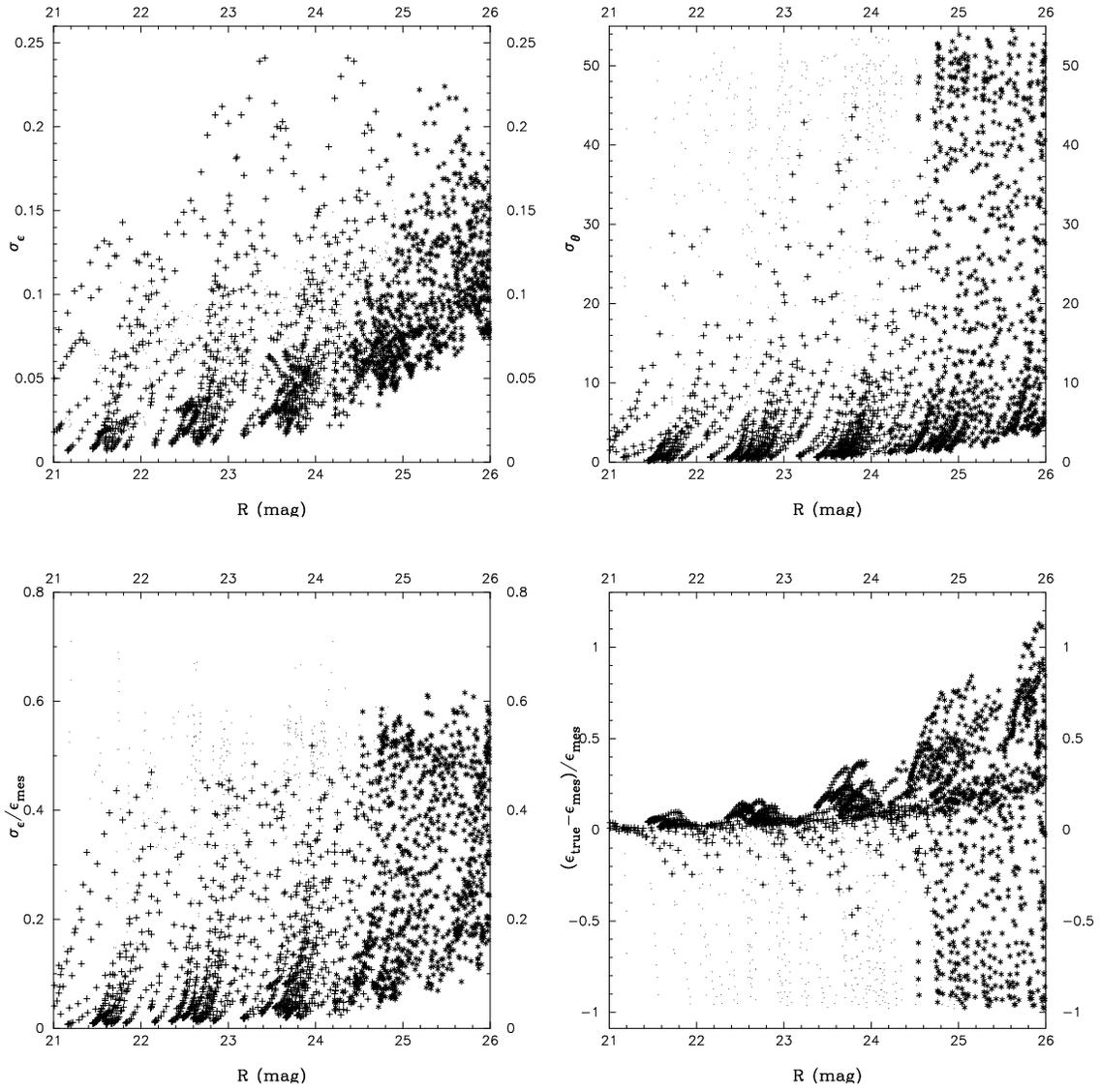

Figure 10

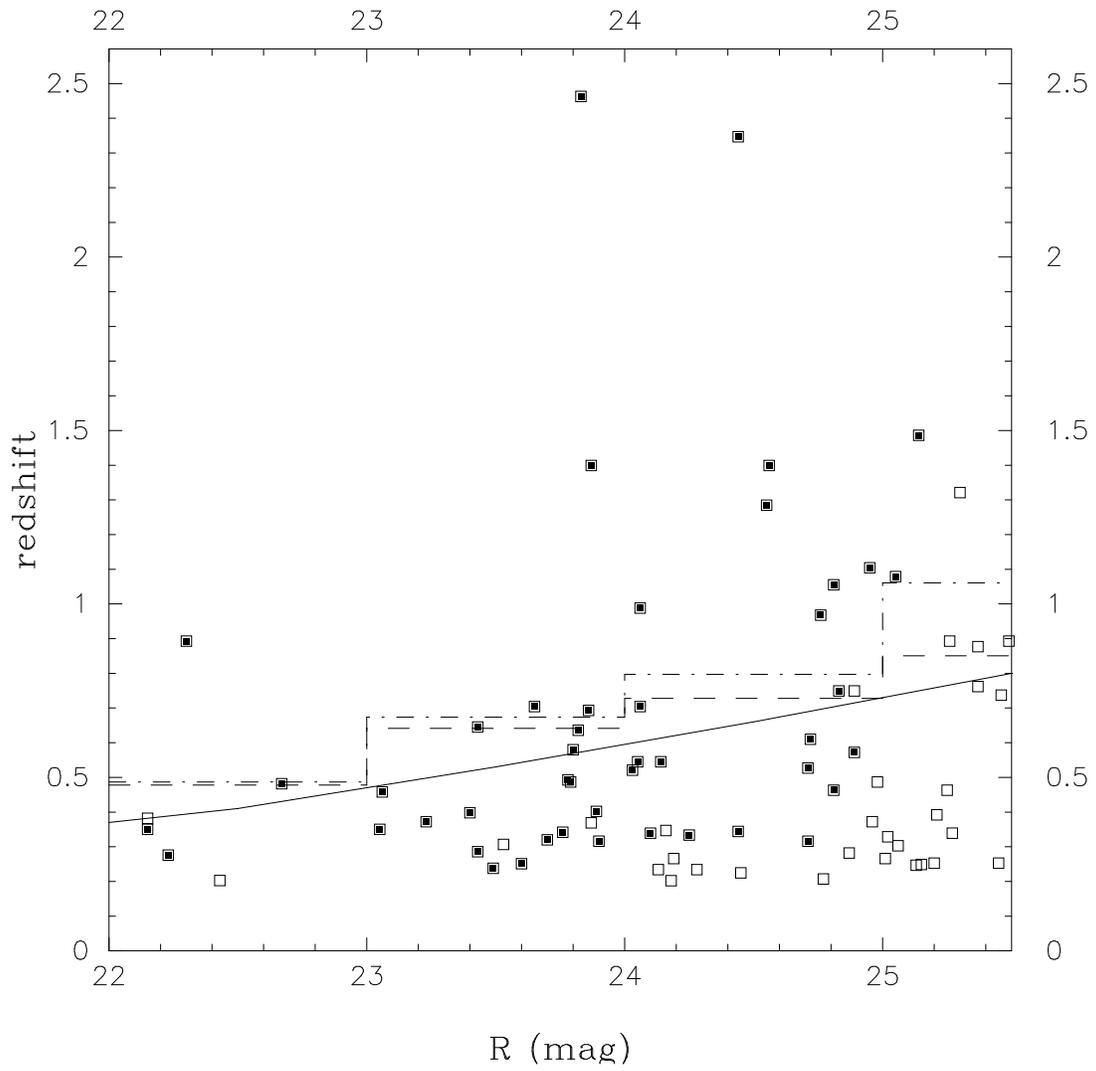

Figure 11